\documentclass[conference]{IEEEtran}
\IEEEoverridecommandlockouts
\usepackage{amsmath,amssymb,amsfonts}
\usepackage{algorithmic}
\usepackage{graphicx}
\usepackage{textcomp}
\usepackage{xcolor}
\usepackage{hyperref}
\usepackage{booktabs}
\usepackage[switch]{lineno} 
\usepackage{etoolbox}

\usepackage[square,numbers]{natbib}

\begin{document}

\title{Insight Into SEER\\
}

\author{
\IEEEauthorblockN{Kasra Lekan}
\IEEEauthorblockA{\textit{Computer Science Department} \\
\textit{University of Virginia}\\
Charlottesville, VA, USA \\
kl5sq@virginia.edu}
\and
\IEEEauthorblockN{Nicki Choquette}
\IEEEauthorblockA{\textit{Computer Science Department} \\
\textit{University of Virginia}\\
Charlottesville, VA, USA \\
nc2uzv@virginia.edu}
}

\maketitle

\footnote{This work was completed on May 8, 2023}\begin{abstract}
Developing test oracles can be inefficient: developer generative oracles are time-intensive and thus costly while automatic oracle generation in the form of regression or exception oracles assumes that the underlying code is correct. To mitigate the high cost of testing oracles, the SEER tool was developed to predict test outcomes without needing assertion statements. The creators of SEER introduced the tool with an overall accuracy of 93\%, precision of 86\%, recall of 94\%, and an F1 score of 90\%. If these results are replicable on new data with perturbations, i.e. SEER is generalizable and robust, the model would represent a significant advancement in the field of automated testing. Consequently, we conducted a comprehensive reproduction of SEER and attempted to verify the model's results on a new dataset.

\end{abstract}

\section{Introduction}

Automatic software testing is an essential aspect of modern software development, as it ensures the reliability and robustness of software applications. The rapid growth of software complexity and the increasing need for continuous integration and delivery have created a demand for more efficient and accurate testing methodologies. One promising approach to address this challenge is the application of neural networks in the domain of automatic software testing. Neural networks can learn to recognize patterns and relationships in code, thereby enabling them to predict potential defects, generate test cases, and even identify the expected behavior of a given software module \cite{dinella_toga_2022}.

\begin{figure}
    \centering
    \includegraphics[width=\linewidth]{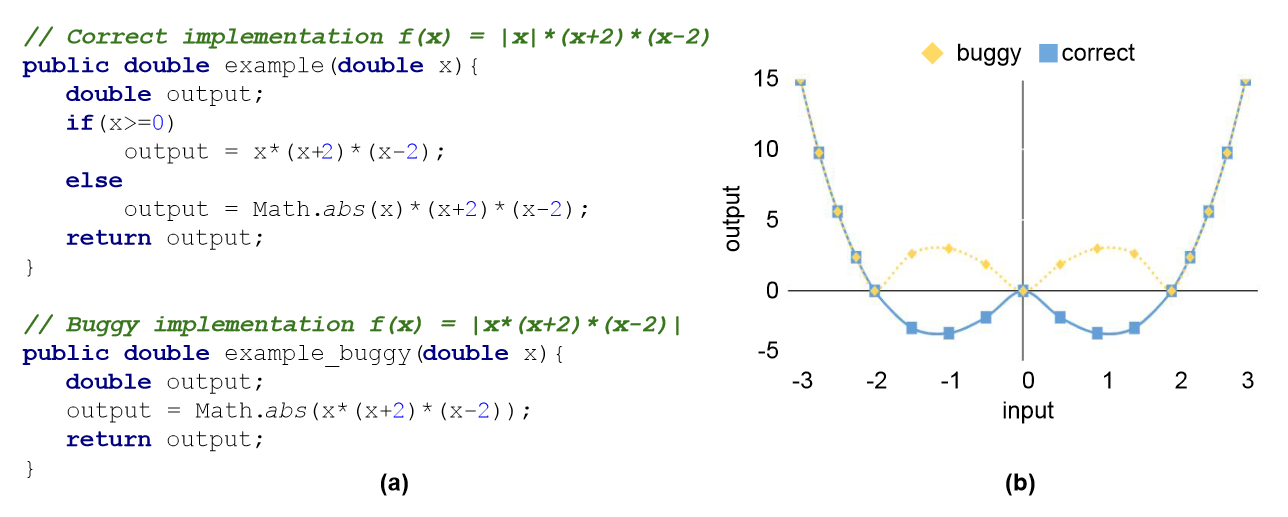}
    \caption{Example of Correct and Buggy MUT with inputs and outputs \cite{ibrahimzada_perfect_2022}}
    \label{fig:seer_example}
\end{figure}

A critical challenge in automatic software testing is the ``oracle problem," which refers to the difficulty of determining the correct output or behavior for a given set of input conditions in a test case. In other words, the oracle problem deals with deciding whether a test has passed or failed, given the observed behavior of the software under test. Traditionally, this has been addressed by using human-written assertions, metamorphic properties, or regression testing. However, these approaches can be time-consuming, error-prone, and often fail to cover the complete set of inputs, e.g. testing between -2 and 2 in Figure \ref{fig:seer_example}.

The SEER model, introduced in the paper \href{https://dl.acm.org/doi/pdf/10.1145/3540250.3549086?casa_token=qBiInpwefF4AAAAA:Sj1vwvJEQuItkPmDcAGHSn-8Jv1ZKjSq6x8odgcmZvaZcbG8bJeWdEUtkirwvLPbyucKoktKrEzC}{Perfect Is the Enemy of Test Oracle,} is a neural network-based approach to automatic software testing that aims to predict the outcomes of test cases without the need for assertion statements. SEER leverages joint embedding of code and tests (Phase 1) and a classifier (Phase 2) to distinguish correct and buggy methods without using assertions. The model is trained on data from Defects4J \cite{just_defects4j_2014} which includes test cases, and correct and buggy versions of the corresponding method under test (MUT). SEER is the first model of its kind, with its ability to predict whether a test will pass or fail with no assertions and no requirement to execute the tests. 

The creators of SEER reported promising results for their model, with an overall accuracy of 93\%, precision of 86\%, recall of 94\%, and an F1 score of 90\%. These results, if replicable and generalizable, indicate that SEER could significantly improve the efficiency and effectiveness of software testing by reducing or removing the reliance on oracles.

Our key contributions are (RQ1) replicating the original SEER results, (RQ2) reproducing the SEER experiment on a new dataset, (RQ3) evaluating the robustness of the SEER model, and (RQ4) determining if SEER uses meaningful input tokens to make predictions via an attention analysis.

\section{Approach}
\noindent High-level approach:
\renewcommand{\labelenumi}{\Alph{enumi}}
\begin{enumerate}
    \item Reproduce SEER's results \cite{ibrahimzada_perfect_2022} on the original dataset.
    \item Transform the new dataset to be compatible with SEER.
    \item Test the pre-trained model on a new dataset.
    \item Evaluate the robustness of SEER to perturbations of the new dataset.
    \item Conduct an explainability analysis.
\end{enumerate}
\renewcommand{\labelenumi}{\arabic{enumi}}

\subsection{Reproduction with Original Dataset}
We utilized scripts provided by the authors to reproduce the headline results. Since the authors reported only recall and precision metrics on the ``unseen'' data, we carried out a more in-depth analysis of the model's performance on that data.

\subsection{New Dataset}

\begin{figure}[h]
    \centering
    \includegraphics[width=.50\linewidth]{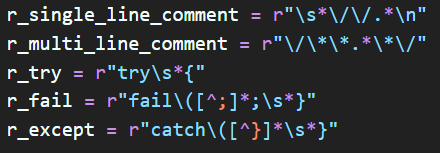}
    \caption{Regex used to replace comments and try-catch blocks.}
    \label{fig:regex}
\end{figure}

The new dataset \cite{10.1145/3611643.3616265} is a dataset comprising real-world MUT-test pairs from 25 Java systems. It encompasses 169,099 test cases (15k+ exception oracles), generated by EvoSuite and, therefore, all passing. Four projects appear in both the SEER training data and the new data (Jsoup, Lang, Time, Collections). However, as demonstrated in Table \ref{tab:common_unique_MUT}, there are considerable differences between the code in these projects across the datasets. These disparities primarily arise because SEER's data (sourced from Defects4J \cite{just_defects4j_2014}) only includes MUTs identified as having bugs, while the new dataset covers a broader range of MUTs from bug-free versions of each project.

\begin{figure}[h]
    \centering
    \includegraphics[width=\linewidth]{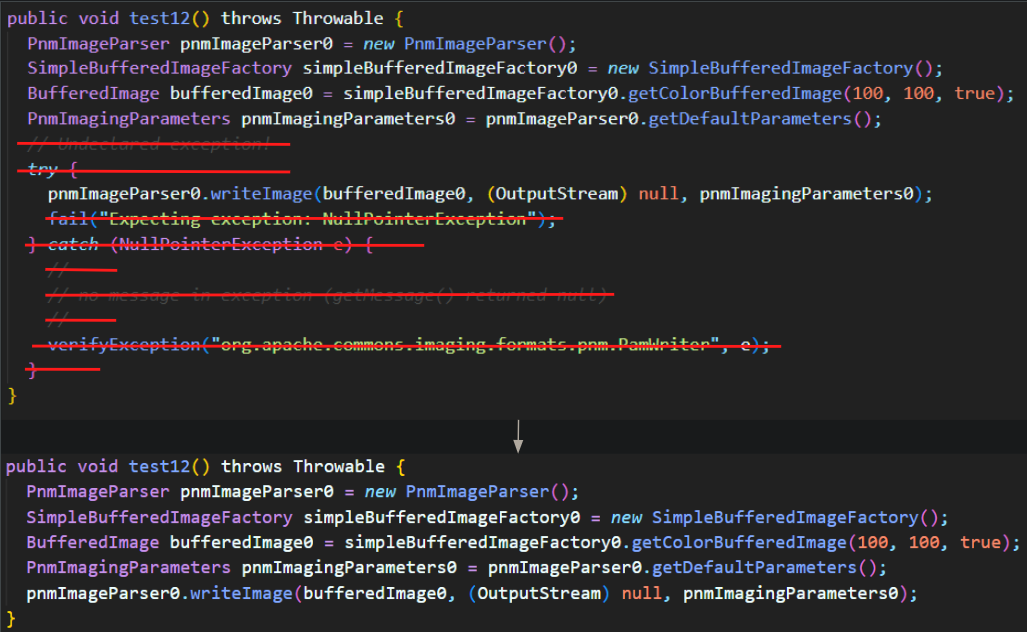}
    \caption{Creating a Failing test from a Passing test by removing a Try-catch block.}
    \label{fig:try-catch}
\end{figure}

To generate data with a ``Fail'' label, we employed REGEX (Figure \ref{fig:regex}) to remove the try-catch blocks from exception oracles, as depicted in Figure \ref{fig:try-catch}, which results in an uncaught exception. Since SEER's training data lacked comments, we applied additional REGEX to remove any comments from the code.

After modifying the data, we attempted to vectorize it for use with SEER. The reproduction package's built-in function was exclusively designed to vectorize data formatted for both Phase 1 and Phase 2 (the new dataset only conforms to the Phase 2 schema). Moreover, the function would throw an exception if any vocab tokens were missing. SEER's failure to recognize new vocabulary is the principal limitation of the SEER model. Consequently, we developed our own vectorization script that fills in missing vocabulary with zeros (equivalent to the padding token) in the vector. During this step, we collected data on which portions of the code and test were out of vocab (Table \ref{tab:stats_all}).

\begin{table}
\centering
\caption{Unique Methods Under Test in projects shared between both datasets.}
\label{tab:common_unique_MUT}
\begin{tabular}{p{1.5cm}p{1.5cm}p{1.5cm}r}
\toprule
        SEER Projects &  SEER Unique MUT (\#) &  New Data Unique MUT (\#) & New Projects \\
\midrule
       Jsoup &         254 &          292 &                 jsoup \\
        Lang &         300 &          995 &         commons-lang3 \\
        Time  &         105 &         1479 &             joda-time\\
 Collections &           3 &          185 &  commons-collections4 \\
\bottomrule
\end{tabular}
\end{table}

\subsection{Testing SEER on New Data}
We partitioned the data by source project and processed it using the classifier. We calculated accuracy (overall and within each class) and F1 score. Then, we compared the results to a weighted coin (a Bernoulli variable with a probability equal to the rate of ``Pass'' labels within the dataset). We selected the weighted coin as a baseline because it represents a purely statistical (``guessing'') classifier, whereas the SEER model should possess a learned semantic understanding of the MUT-test pairs. Since some projects had relatively small samples, we repeated the coin flip at least 10,000 times and averaged the results. We calculated three key metrics for evaluating the model's performance against the baseline: the change in accuracy (Accuracy $\Delta$), F1 score (F1 $\Delta$), and Failure class accuracy (Fail Accuracy $\Delta$). Considering that most production tests pass in any given code base and regression testing always generates passing tests, we believe that evaluating SEER's performance on the failing cases in the dataset is the most revealing approach. Therefore, we primarily focused on Fail Accuracy $\Delta$ during our evaluation.

Utilizing the data collected on total out-of-vocab tokens, we also performed an identical analysis while retaining only cases where the percentage of missing tokens was less than 50\%, 25\%, 20\%, 15\%, 10\%, and 5\%. We hoped this would enable us to approximate how well SEER would perform if the model was enhanced to handle unseen vocabulary. As the threshold for the percentage of missing tokens became more restrictive, some projects were excluded from the data, and some were left with too small a sample to confidently draw conclusions. To avoid reporting small-sample anomalies, we only used project data where $N>20$. Thus, we tended to rely on data from all projects combined or from the average across projects rather than judge on a project-by-project basis once we crossed the 15\% threshold. We compared each subset of the data to the superset results via our three key metrics, e.g. the difference in Accuracy $\Delta$ with a subset of data and Accuracy $\Delta$ with all data.

\subsection{Robustness Testing}
Robustness testing is based on research in adversarial learning, specifically using perturbations of code that do not meaningfully alter the state of the underlying program \cite{yefet_adversarial_2020}. Due to SEER's vocabulary limitations, our options for robustness testing were quite limited. For example, if the variable ``foo'' was in the model's vocabulary and we wanted to change it to ``bar,'' which was not in the model's vocabulary, the vectorization of the variable would be 0. Thus, we opted to conduct an analysis with comments in the code using the SEER vocabulary. In addition to our primary dataset, which was stripped of comments as described in the New Dataset section, we tested four scenarios: (1) Comments in the new dataset were preserved. Given the vocabulary limitations discussed earlier, we found that many preserved comments were out-of-vocab and replaced with 0s in the vectorized data. (2) Comments were added at the end of every test, declaring that an exception would occur (``// report (Exception) diagnose problems debugging might be helpful''). (3) The comment was added at the end of every MUT. (4) The comment was added at the end of both the MUT and the test case.

\subsection{Explainability Analysis}
\begin{figure}[h]
    \centering
    \includegraphics[width=.8\linewidth]{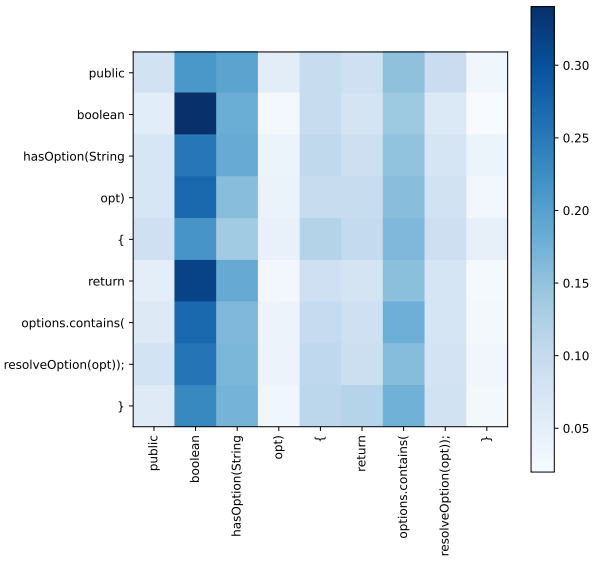}
    \caption{Example heat map version of a SA matrix from Phase 2 unseen data.}
    \label{fig:14915}
\end{figure}

In order to evaluate why and how SEER predicts a certain label for a MUT-test pair, it was necessary to conduct an explainability analysis. During the training phase of SEER, Self Attention (SA) is utilized to assess the relative significance of a token within a statement \cite{ibrahimzada_perfect_2022}. Self Attention outputs a matrix with each entry $w_{i_j}$ signifying the importance of token $i$ given the token at location $j$ (Figure \ref{fig:14915}). When a predicted MUT-test label is ``Fail", the statements for which a threshold percentage (we set our threshold to 5\%) of tokens overlap with the attended tokens are meant to indicate the buggy statements.



In our attention analysis, we encountered several challenges, one of which was the discrepancy between the main.java file's implementation and the algorithm described in the original paper. Specifically, main.java required not only the MUT and the SA matrix but also a list of differences between the correct MUT and a buggy version of the MUT. This approach undermines the explainability of the SEER model, as having both the correct and buggy versions of the code implies that the correct code is already known. Thus, we modified main.java to parse the code statements and utilize them in place of the previously required buggy differences.

Our analysis of the self-attention results comprised the following steps:
(1) We executed attention\_analysis.py to generate the self-attention matrices for Phase 2 unseen data and the new data.
(2) We then ran main.java on the attention matrices for the aforementioned data and output the results to a text file.
(3) Finally, we examined instances where SEER failed to predict the correct label to identify the most attended statements for SEER and determine the reasons behind its incorrect predictions.

\section{Evaluation}

\begin{table}
\centering
\caption{Performance of SEER on New Data (no comments) with varying maximum \% of tokens out-of-vocab. $\delta = \Delta(threshold\_data) - \Delta(all\_data)$}
\label{tab:vocab_analysis}
\begin{tabular}{p{2cm}lp{1cm}p{1cm}p{1cm}}
\toprule
Max \% Missing Vocabulary &       N &  Fail Accuracy $\delta$ &  Accuracy $\delta$ &  F1 $\delta$ \\
\midrule
                         50\% &  145474 &                -0.000 &           -0.010 &     -0.005 \\
                         25\% &   35918 &                -0.016 &            0.011 &      0.006 \\
                         20\% &   10156 &                -0.050 &           -0.007 &     -0.002 \\
                         15\% &    2643 &                -0.022 &           -0.074 &     -0.045 \\
\bottomrule
\end{tabular}
\end{table}

\subsection{RQ1: Can the results of the SEER paper be reproduced?}
Aside from regenerating the SEER training data from Defects4J and training the models, we reproduced all the results from the original paper. The Phase 2 data used by the authors in their headline results and training (``whole'') contained many large projects. Another dataset (``unseen'') which consisted of some smaller projects was used to demonstrate generalizability as SEER's precision and recall were about 10\% lower than the scores on ``whole.'' The authors argued that this inferior performance was due to an out-of-distribution problem associated with the ``unseen'' data compared to ``whole'' data. 

Upon reproducing the results on the ``unseen'' data, we obtained an accuracy of 66.7\% (with 81.6\% pass accuracy and 14.1\% fail accuracy). The model's performance on the ``unseen'' data was not only almost 30\% lower than ``whole,'' but the results were also biased with respect to class.

\subsection{RQ2: To what extent does the SEER model generalize to new data?}
\subsubsection{Overall Results}
The overall dataset is imbalanced with a pass rate of 91.1\% and the data was missing 35\% of tokens. With an overall accuracy of 89.2\% (with 97.3\% pass accuracy and 7.0\% fail accuracy) and an F1 score of 94.3\%, the model performs better than we expected on the new data, especially considering that 35\% of tokens are missing. Distinct from the results of the original paper, though, the model is severely biased with respect to class. Compared to the baseline of a weighted coin, the model had a 5.6\% increase in accuracy, a 3.3\% increase in F1-score, and a 2.0\% decrease in accuracy for ``Fail'' labels.

\subsubsection{Project Results}
Performance varies considerably across projects (Table \ref{tab:results_all}). Fail Accuracy $\Delta$s range from 16.5\% (commons-pool2) to -48.1\% (scribejava) with six projects showing improvement and 19 experiencing a decrease in Fail Accuracy over a weighted coin. Intriguingly, overall Accuracy $\Delta$s and F1 $\Delta$s tend to be inversely related to Fail Accuracy $\Delta$s. The two projects with the highest fail-class accuracy $\Delta$, commons-pool2 and commons-collections4, are the only ones with negative overall Accuracy $\Delta$s and F1 $\Delta$s. This indicates that the model is overclassifying MUT-test pairs as ``Pass'' and exhibits class bias across all projects.

\subsubsection{Vocabulary-restricted Results}
As illustrated in Table \ref{tab:vocab_analysis}, the model does perform better as the percentage of in-vocab tokens increases, defying our expectations. In terms of overall performance metrics, the model has mixed results for each threshold until 15\%, the most significant impacts are two Failure Accuracy $\Delta$s which decrease 1.6\% and 5.0\% at the 25\% and 20\% thresholds respectively. At 15\%, SEER performs considerably worse with an accuracy delta decrease of 7.4\%. The decline in performance as more tokens are covered is perplexing. It might be an anomaly due to the significantly smaller amount of data in the later samples.

Subsets with a percentage of missing tokens less than 10\% and 5\% had samples too small (N=473, N=131) for meaningful evaluation. A more granular analysis by project is impossible due to the sparsity of the data at high levels of vocab match.

\begin{table}[h]
\centering
\caption{Performance of SEER on New Data with different comment types, compared to a no comment baseline. $\delta = SEER\_val(comment_type) - SEER\_val(no\_comments)$, e.g. the ``No Comments`` row is all zeros because its values are subtracted by themselves.}
\label{tab:comment_analysis}
\begin{tabular}{lp{1cm}p{1cm}p{1cm}}
\toprule
            Comment Type &  Fail Accuracy $\delta$ &  Accuracy $\delta$ &  F1 $\delta$ \\
\midrule
             No Comments &                0.0000 &           0.0000 &     0.0000 \\
      Preserved Comments &                0.0055 &          -0.0002 &    -0.0002 \\
      Added Test Comment &               -0.0086 &           0.0005 &     0.0003 \\
       Added MUT Comment &               -0.0446 &           0.0096 &     0.0055 \\
 Added MUT/Test Comments &               -0.0558 &           0.0135 &     0.0077 \\
\bottomrule
\end{tabular}
\end{table}

\subsection{RQ3: To what extent is the SEER model robust to semantically irrelevant tokens in input?}
Table \ref{tab:comment_analysis} summarizes the performance differences between different types of comments. Preserving the comments in the tests and code has no significant effect on accuracy and F1 scores, while increasing overall failing accuracy by .55\%. This is likely due to the out-of-vocab words used in some comments in the dataset. When we add comments that are in-vocab the effect is more pronounced, e.g. Listing \ref{listing:robust_misclassify}. When adding the same comment at the end of code and tests, the overall failure accuracy falls by 5.6\% and the overall accuracy rises by 1.4\%. It is worth noting that the SEER model was not trained on commented code or tests. As a result, the model likely considered the added tokens for each comment as code to be executed rather than irrelevant tokens. This partly explains why the effect size was more pronounced when adding comments to code compared to adding comments to tests. Code in the new dataset was generally shorter than tests so adding a few tokens likely had a more pronounced impact if the model was interpreting them as meaningful input.

\subsection{RQ4: To what extent does the SEER model use meaningful input tokens to make its predictions?}
To determine if SEER is using meaningful input we consider two cases: when SEER incorrectly predicts ``Pass" or incorrectly predicts ``Fail". 

\begin{figure}[h]
    \centering
    \includegraphics[width=.8\linewidth]{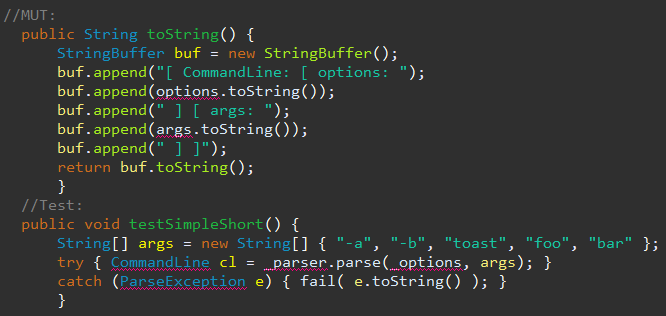}
    \caption{Example where SEER incorrectly predicted ``Pass" from Phase 2 unseen data.}
    \label{fig:14984}
\end{figure}

Firstly, we have a Phase 2 unseen example of a MUT-test pair where SEER incorrectly predicted ``Pass" (Figure \ref{fig:14984}). Here, the two statements with a contribution percent $>0$ are assignments to `valueList', which makes sense as both of the assignments depend on parameters, only one of which is passed in. This greatly calls into question the ability of SEER to correctly predict test outcomes, given that SEER was giving the most attention to the buggy parts of the code and still failed to predict correctly.

\begin{figure}[h]
    \centering
    \includegraphics[width=.8\linewidth]{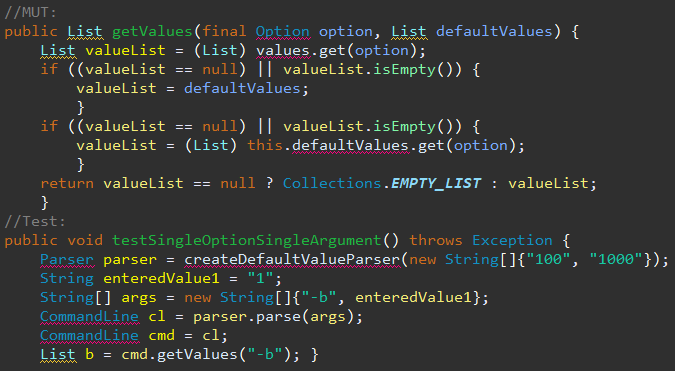}
    \caption{Example where SEER incorrectly predicted ``Fail" from Phase 2 unseen data.}
    \label{fig:15160}
\end{figure}

Secondly, we have a Phase 2 unseen example of a MUT-test pair where SEER incorrectly predicted ``Fail" (Figure \ref{fig:15160}). Of note, it makes sense that this test will pass, due to its utilization of a try-catch block. However, the statement identified by SEER as having the highest contribution percent was `buf.append(args.toString());', which is not buggy. Additonally, we noticed a pattern of SEER incorrectly predicting ``Fail" when tests contained a try-catch block which occurred in 69.7\% of cases. Having a try-catch block should not have a correlation with fail prediction necessarily.

As demonstrated, at times SEER can identify meaningful tokens of the MUT. However, even when it correctly identifies these tokens, it is inconsistent in its ability to make a correct prediction. This demonstrates a pitfall of SEER and calls into question how well the model would perform. Without clear reasoning behind its predictions, it is difficult to have confidence in SEER.

\section{Conclusions}
In summary, our investigation reveals significant limitations in the SEER model, despite its novelty and impressive performance. We have identified four key limitations that need to be addressed for further advancements in this area.

Firstly, the model's inability to generalize to new vocabulary hinders its applicability to arbitrary MUT-test pairs. To overcome this limitation, we recommend implementing byte-pair encoding. We observed a slight decrease in SEER's performance on data subsets with a higher percentage of tokens in the vocabulary. Ultimately, this area needs further investigation.

Secondly, our analysis of the new dataset demonstrated significant variations in the SEER model's performance across different projects. Although the model achieved an impressive overall accuracy improvement of 5.6\% over baseline, it exhibited severe bias towards the ``Pass'' class. These findings align with reproduction of the model's results on ``unseen'' data from the SEER paper, where the model's performance deteriorated and revealed bias when not trained on the underlying data. Interestingly, projects such as Jsoup, Lang, Time, and Collections, which were shared between the original training data and new data, exhibited higher fail-class accuracy $\Delta$s. This highlights the need for further investigation and mitigation of biases within the SEER model.

Thirdly, we identified that the model was fairly robust to added comments within the tests but had mixed results when comments were added to the code with a 5.6\% decrease in failure class accuracy but a 1.4\% increase in overall accuracy.

Finally, SEER failed to produce adequately explainable results. Case studies at large software companies such as Facebook \cite{distefano_scaling_2019} have shown that if developers lose faith in a tool, it will fall to disuse. Much like how in medical settings we have come to expect neural network models to be explainable in order to verify the results prior to application, developers need to understand and trust SEER's classification in order for it to be deployed. To that point, we did not find sufficient evidence in the explainability analysis to ``trust'' SEER.

Moving forward, it is crucial to modify the SEER model to incorporate byte-pair encoding and retrain it accordingly. This modification is expected to enhance the accuracy and F1 scores for all projects in the new dataset. However, we acknowledge that this step alone may not fully resolve the underlying issues related to generalizability. Therefore, future research should focus on conducting a robust evaluation that includes more advanced manipulations of the underlying MUTs or tests. By expanding the evaluation to encompass diverse scenarios, we can gain deeper insights into the model's capabilities and limitations.

In conclusion, the SEER model shows promise but requires significant refinements to overcome the limitations identified in this study. By addressing these limitations and conducting further investigations, we can pave the way for advancements in automated software testing and analysis.

\section*{Acknowledgment}

We would like to thank Soneya Binta Hossain prompting for this replication study and providing the new dataset in addition to her continued support throughout its completion. Additionally, we thank Professor Sebastian Elbaum for his support and suggestions for improving our research methodology. Finally, we would like to thank the authors of the SEER paper for answering our questions about their replication package.


\begin{thebibliography}{6}
\providecommand{\natexlab}[1]{#1}
\providecommand{\url}[1]{\texttt{#1}}
\expandafter\ifx\csname urlstyle\endcsname\relax
  \providecommand{\doi}[1]{doi: #1}\else
  \providecommand{\doi}{doi: \begingroup \urlstyle{rm}\Url}\fi

\bibitem[Dinella et~al.(2022)Dinella, Ryan, Mytkowicz, and
  Lahiri]{dinella_toga_2022}
E.~Dinella, G.~Ryan, T.~Mytkowicz, and S.~K. Lahiri.
\newblock {TOGA}: a neural method for test oracle generation.
\newblock In \emph{Proceedings of the 44th {International} {Conference} on
  {Software} {Engineering}}, pages 2130--2141, Pittsburgh Pennsylvania, May
  2022. ACM.
\newblock ISBN 978-1-4503-9221-1.
\newblock \doi{10.1145/3510003.3510141}.
\newblock URL \url{https://dl.acm.org/doi/10.1145/3510003.3510141}.

\bibitem[Distefano et~al.(2019)Distefano, Fähndrich, Logozzo, and
  O'Hearn]{distefano_scaling_2019}
D.~Distefano, M.~Fähndrich, F.~Logozzo, and P.~W. O'Hearn.
\newblock Scaling static analyses at {Facebook}.
\newblock \emph{Communications of the ACM}, 62\penalty0 (8):\penalty0 62--70,
  July 2019.
\newblock ISSN 0001-0782, 1557-7317.
\newblock \doi{10.1145/3338112}.
\newblock URL \url{https://dl.acm.org/doi/10.1145/3338112}.

\bibitem[Hossain et~al.(2023)Hossain, Filieri, Dwyer, Elbaum, and
  Visser]{10.1145/3611643.3616265}
S.~B. Hossain, A.~Filieri, M.~B. Dwyer, S.~Elbaum, and W.~Visser.
\newblock Neural-based test oracle generation: A large-scale evaluation and
  lessons learned.
\newblock In \emph{Proceedings of the 31st ACM Joint European Software
  Engineering Conference and Symposium on the Foundations of Software
  Engineering}, ESEC/FSE 2023, page 120–132, New York, NY, USA, 2023.
  Association for Computing Machinery.
\newblock ISBN 9798400703270.
\newblock \doi{10.1145/3611643.3616265}.
\newblock URL \url{https://doi.org/10.1145/3611643.3616265}.

\bibitem[Ibrahimzada et~al.(2022)Ibrahimzada, Varli, Tekinoglu, and
  Jabbarvand]{ibrahimzada_perfect_2022}
A.~R. Ibrahimzada, Y.~Varli, D.~Tekinoglu, and R.~Jabbarvand.
\newblock Perfect is the enemy of test oracle.
\newblock In \emph{Proceedings of the 30th {ACM} {Joint} {European} {Software}
  {Engineering} {Conference} and {Symposium} on the {Foundations} of {Software}
  {Engineering}}, pages 70--81, Singapore Singapore, Nov. 2022. ACM.
\newblock ISBN 978-1-4503-9413-0.
\newblock \doi{10.1145/3540250.3549086}.
\newblock URL \url{https://dl.acm.org/doi/10.1145/3540250.3549086}.

\bibitem[Just et~al.(2014)Just, Jalali, and Ernst]{just_defects4j_2014}
R.~Just, D.~Jalali, and M.~D. Ernst.
\newblock {Defects4J}: a database of existing faults to enable controlled
  testing studies for {Java} programs.
\newblock In \emph{Proceedings of the 2014 {International} {Symposium} on
  {Software} {Testing} and {Analysis}}, pages 437--440, San Jose CA USA, July
  2014. ACM.
\newblock ISBN 978-1-4503-2645-2.
\newblock \doi{10.1145/2610384.2628055}.
\newblock URL \url{https://dl.acm.org/doi/10.1145/2610384.2628055}.

\bibitem[Yefet et~al.(2020)Yefet, Alon, and Yahav]{yefet_adversarial_2020}
N.~Yefet, U.~Alon, and E.~Yahav.
\newblock Adversarial {Examples} for {Models} of {Code}, Oct. 2020.
\newblock URL \url{http://arxiv.org/abs/1910.07517}.
\newblock arXiv:1910.07517 [cs].

\end{thebibliography}

\section*{Appendix}

Full project names in New Data: commons-jexl3-3.2.1-src, commons-weaver-2.0-src, commons-geometry-1.0-src, commons-rng-1.4-src, springside4, commons-beanutils-1.9.4, async-http-client, commons-configuration2-2.8.0-src, commons-vfs-2.9.0, commons-net-3.8.0, commons-validator-1.7, commons-lang3-3.12.0-src, commons-jcs3-3.1-src, spark, bcel-6.5.0-src, commons-numbers-1.0-src, scribejava, commons-collections4-4.4-src, joda-time, commons-imaging-1.0-alpha3-src, commons-dbutils-1.7, commons-pool2-2.11.1-src, jsoup, JSON-java, http-request.

\begin{table*}[h!]
\centering
\caption{New Dataset Statistics (no comments). Pass/Fail rates demonstrate how often a class is present in the data, e.g. a balanced dataset would have .5 for both columns. Missing Tokens refer to tokens that the model is unable to encode. Tokens are generated by running Python's string function split() which splits strings by spaces.}
\label{tab:stats_all}
\begin{tabular}{|l|r|p{1cm}|p{1cm}|p{1cm}|p{1cm}|p{1cm}|}
\toprule
                project &       N &  Dataset Pass Rate &  Dataset Fail Rate &  Missing MUT Token Rate &  Missing Test Token Rate &  Missing Overall Token Rate \\
\midrule
          commons-pool2 &   11244 &                 0.9924 &                 0.0076 &                    0.41 &                     0.54 &                        0.51 \\
   commons-collections4 &    1389 &                 0.8762 &                 0.1238 &                    0.45 &                     0.46 &                        0.46 \\
        commons-numbers &   39866 &                 0.9938 &                 0.0062 &                    0.16 &                     0.29 &                        0.26 \\
              JSON-java &   12911 &                 0.9844 &                 0.0156 &                    0.23 &                     0.51 &                        0.44 \\
                  spark &    5280 &                 0.9256 &                 0.0744 &                    0.36 &                     0.51 &                        0.46 \\
              joda-time &   27480 &                 0.9355 &                 0.0645 &                    0.29 &                     0.49 &                        0.44 \\
           http-request &    4069 &                 0.9956 &                 0.0044 &                    0.27 &                     0.44 &                        0.36 \\
                  jsoup &    8002 &                 0.9353 &                 0.0647 &                    0.08 &                     0.43 &                        0.28 \\
          commons-lang3 &   12118 &                 0.8818 &                 0.1182 &                    0.20 &                     0.35 &                        0.27 \\
                   bcel &   15379 &                 0.8318 &                 0.1682 &                    0.28 &                     0.45 &                        0.36 \\
            commons-rng &    1641 &                 0.7367 &                 0.2633 &                    0.28 &                     0.40 &                        0.34 \\
      commons-validator &    2616 &                 0.8498 &                 0.1502 &                    0.28 &                     0.37 &                        0.32 \\
           commons-jcs3 &    4562 &                 0.8637 &                 0.1363 &                    0.32 &                     0.43 &                        0.39 \\
       commons-geometry &    5438 &                 0.7696 &                 0.2304 &                    0.33 &                     0.47 &                        0.41 \\
        commons-imaging &    2858 &                 0.7978 &                 0.2022 &                    0.30 &                     0.47 &                        0.41 \\
 commons-configuration2 &    1253 &                 0.6744 &                 0.3256 &                    0.28 &                     0.44 &                        0.35 \\
        commons-dbutils &     731 &                 0.7893 &                 0.2107 &                    0.31 &                     0.48 &                        0.43 \\
            commons-vfs &    1072 &                 0.7276 &                 0.2724 &                    0.33 &                     0.45 &                        0.40 \\
          commons-jexl3 &    2946 &                 0.7634 &                 0.2366 &                    0.27 &                     0.45 &                        0.35 \\
            springside4 &    2590 &                 0.7158 &                 0.2842 &                    0.25 &                     0.36 &                        0.31 \\
      async-http-client &     163 &                 0.7117 &                 0.2883 &                    0.38 &                     0.49 &                        0.44 \\
         commons-weaver &     232 &                 0.6638 &                 0.3362 &                    0.34 &                     0.45 &                        0.39 \\
            commons-net &    3369 &                 0.6628 &                 0.3372 &                    0.32 &                     0.35 &                        0.34 \\
      commons-beanutils &    1708 &                 0.5872 &                 0.4128 &                    0.24 &                     0.40 &                        0.30 \\
             scribejava &     182 &                 0.5110 &                 0.4890 &                    0.32 &                     0.46 &                        0.39 \\
                    all &  169099 &                 0.9110 &                 0.0890 &                    0.24 &                     0.41 &                        0.35 \\
\bottomrule
\end{tabular}
\end{table*}

\begin{table*}[h!]
\centering
\caption{SEER Results on New Data (no comments), sorted by failure accuracy $\Delta$. \\ $\Delta = seer\_val - coin\_val$}
\label{tab:results_all}
\resizebox{\textwidth}{!}{
\begin{tabular}{|l|p{1cm}|p{1cm}|p{1cm}|p{1cm}|p{1cm}|p{1cm}|p{1cm}|p{1cm}|r|r|r|r|}
\toprule
                project &  Fail Accuracy $\Delta$ &  Accuracy $\Delta$ &  F1 $\Delta$ &  Accuracy &  Pass Class Accuracy &  Fail Class Accuracy &      F1 &  Coin Accuracy &      tp &    fn &    tn &     fp \\
\midrule
          commons-pool2 &                    0.1647 &              -0.0611 &    -0.0321 &    0.9238 &               0.9294 &               0.1882 &  0.9603 &         0.9849 &   10371 &   788 &    16 &     69 \\
   commons-collections4 &                    0.0574 &              -0.3932 &    -0.3289 &    0.3909 &               0.4215 &               0.1744 &  0.5481 &         0.7841 &     513 &   704 &    30 &    142 \\
        commons-numbers &                    0.0482 &               0.0055 &     0.0028 &    0.9930 &               0.9988 &               0.0562 &  0.9965 &         0.9875 &   39571 &    46 &    14 &    235 \\
              JSON-java &                    0.0199 &               0.0149 &     0.0076 &    0.9830 &               0.9982 &               0.0249 &  0.9914 &         0.9681 &   12687 &    23 &     5 &    196 \\
                  spark &                    0.0114 &               0.0517 &     0.0292 &    0.9112 &               0.9783 &               0.0763 &  0.9532 &         0.8595 &    4781 &   106 &    30 &    363 \\
              joda-time &                    0.0096 &               0.0475 &     0.0264 &    0.9291 &               0.9877 &               0.0796 &  0.9631 &         0.8816 &   25391 &   317 &   141 &   1631 \\
           http-request &                   -0.0185 &               0.0017 &     0.0008 &    0.9931 &               0.9975 &               0.0000 &  0.9965 &         0.9914 &    4041 &    10 &     0 &     18 \\
                  jsoup &                   -0.0290 &               0.0398 &     0.0222 &    0.9199 &               0.9817 &               0.0270 &  0.9582 &         0.8801 &    7347 &   137 &    14 &    504 \\
          commons-lang3 &                   -0.0440 &               0.0619 &     0.0382 &    0.8548 &               0.9614 &               0.0594 &  0.9211 &         0.7929 &   10274 &   412 &    85 &   1347 \\
                   bcel &                   -0.0487 &               0.0901 &     0.0609 &    0.8116 &               0.9520 &               0.1168 &  0.8937 &         0.7215 &   12179 &   614 &   302 &   2284 \\
            commons-rng &                   -0.0817 &               0.0910 &     0.0780 &    0.7087 &               0.8958 &               0.1852 &  0.8192 &         0.6177 &    1083 &   126 &    80 &    352 \\
      commons-validator &                   -0.0827 &               0.0285 &     0.0209 &    0.7733 &               0.8965 &               0.0763 &  0.8705 &         0.7448 &    1993 &   230 &    30 &    363 \\
           commons-jcs3 &                   -0.1115 &               0.0904 &     0.0579 &    0.8549 &               0.9838 &               0.0386 &  0.9213 &         0.7645 &    3876 &    64 &    24 &    598 \\
       commons-geometry &                   -0.1684 &               0.1160 &     0.0929 &    0.7604 &               0.9692 &               0.0630 &  0.8616 &         0.6444 &    4056 &   129 &    79 &   1174 \\
        commons-imaging &                   -0.1860 &               0.1085 &     0.0815 &    0.7883 &               0.9851 &               0.0121 &  0.8813 &         0.6798 &    2246 &    34 &     7 &    571 \\
 commons-configuration2 &                   -0.2117 &               0.1048 &     0.1153 &    0.6608 &               0.9337 &               0.0956 &  0.7878 &         0.5560 &     789 &    56 &    39 &    369 \\
        commons-dbutils &                   -0.2175 &               0.1115 &     0.0864 &    0.7825 &               0.9913 &               0.0000 &  0.8780 &         0.6710 &     572 &     5 &     0 &    154 \\
            commons-vfs &                   -0.2295 &               0.0567 &     0.0643 &    0.6660 &               0.8974 &               0.0479 &  0.7964 &         0.6093 &     700 &    80 &    14 &    278 \\
          commons-jexl3 &                   -0.2324 &               0.1198 &     0.0995 &    0.7600 &               0.9898 &               0.0187 &  0.8630 &         0.6402 &    2226 &    23 &    13 &    684 \\
            springside4 &                   -0.2353 &               0.1078 &     0.1058 &    0.6985 &               0.9569 &               0.0476 &  0.8196 &         0.5907 &    1774 &    80 &    35 &    701 \\
      async-http-client &                   -0.2580 &               0.1017 &     0.1047 &    0.6871 &               0.9569 &               0.0213 &  0.8132 &         0.5854 &     111 &     5 &     1 &     46 \\
         commons-weaver &                   -0.2797 &               0.1098 &     0.1293 &    0.6638 &               0.9740 &               0.0513 &  0.7937 &         0.5540 &     150 &     4 &     4 &     74 \\
            commons-net &                   -0.3013 &               0.1051 &     0.1281 &    0.6581 &               0.9736 &               0.0379 &  0.7905 &         0.5530 &    2174 &    59 &    43 &   1093 \\
      commons-beanutils &                   -0.3577 &               0.0067 &     0.0873 &    0.5299 &               0.8584 &               0.0624 &  0.6820 &         0.5232 &     861 &   142 &    44 &    661 \\
             scribejava &                   -0.4811 &               0.0068 &     0.1631 &    0.5055 &               0.9785 &               0.0112 &  0.6691 &         0.4987 &      91 &     2 &     1 &     88 \\
                    all &                   -0.0195 &               0.0557 &     0.0325 &    0.8924 &               0.9728 &               0.0699 &  0.9428 &         0.8367 &  149857 &  4196 &  1051 &  13995 \\
\bottomrule
\end{tabular}
}
\end{table*}

\end{document}